\documentclass[prb, fleqn,twocolumn, showpacs, showkeys, floatfix]{revtex4}

\usepackage{graphicx}
\usepackage[dvips]{epsfig}
\usepackage{amsmath,amssymb}

\newcommand\bgi{\begin{itemize}}
\newcommand\egi{\end{itemize}}
\newcommand\bga{\begin{eqnarray*}}
\newcommand\ega{\end{eqnarray*}}
\newcommand\bgan{\begin{eqnarray}}
\newcommand\egan{\end{eqnarray}}

\begin{document}

\title{Spin--dependent transition rates through exchange coupled localized spin pairs during coherent spin excitation}
\author{A. Gliesche$^{a,b}$, C. Michel$^{a,c}$, V. Rajevac$^{d}$, K. Lips$^d$, S.D. Baranovskii$^c$, F. Gebhard$^c$ and C. Boehme$^a$~\footnote{email: boehme@physics.utah.edu}}

\affiliation{$^a$University of Utah, Physics Department, 115S
1400E, Salt Lake City, Utah 84112, USA}

\affiliation{$^b$Institut de th\'eorie des ph\'enom\`enes
physiques, Ecole Polytechnique F\'ed\'erale,  CH-1015
EPF-Lausanne, Switzerland}

\affiliation{$^c$Department of Physics and Material Sciences
Center, Philipps-University, Renthof 5, D-35032 Marburg, Germany}

\affiliation{$^d$Hahn-Meitner-Institut Berlin,
Kekul$\acute{e}$str. 5, D-12489 Berlin, Germany}

\date{\today}

\begin{abstract}
The effect of exchange interactions within spin pairs on
spin--dependent transport and recombination rates through
localized states in semiconductors during coherent electron spin
resonant excitation is studied theoretically. It is shown that for
identical spin systems, significant quantitative differences are
to be expected between the results of pEDMR/pODMR experiments were
permutation symmetry is the observable as compared to pESR
experiments with polarization as the observable. It is predicted
that beat oscillations of the spin nutations and not the nutations
themselves dominate the transport or recombination rates when the
exchange coupling strength or the field strength of the exciting
radiation exceed the difference of the Zeeman energies within the
spin pair. Furthermore, while the intensities of the rate
oscillations decrease with increasing exchange within the spin
pairs, the singlet and triplet signals retain their relative
strength. This means that pEDMR and pODMR experiments could allow
better experimental access to ESR forbidden singlet transitions
which are hardly or not at all accessible with conventional pulsed
electron spin resonance spectroscopy.
\end{abstract}

 \maketitle
\section{Introduction}
In recent years, pulsed electrically and optically detected
magnetic resonance methods (pEDMR and pODMR, respectively) have
attracted increasing attention due to their higher sensitivity in
comparison to the conventional pulsed electron spin resonance
(pESR) technique. While pODMR has been applied to the
investigation of molecular excited states and radical pair
analysis for more than 30 years~\cite{Har:1973,
Glasbeek:1979,oort,moeb1}, pEDMR studies have been reported only
recently when different electrical detection schemes for spin
coherence became available allowing the observation of pulsed
electron~\cite{Boe7,Petta:2005} and even nuclear spin
resonances~\cite{Mach:2003}. For any pODMR and pEDMR experiment
there must be a spin--dependent electronic mechanism which encodes
spin--information into electronic transitions which are then
detected either by means of their radiative emissions (with pODMR)
or through charging and recombination (with pEDMR). The latter has
been studied in the recent past in particular to find potential
spin to charge--conversion mechanisms for the electric readout in
solid state based spin quantum computers\cite{Boe7,Kopp:2006}.

PODMR and pEDMR are typically performed as transient nutation
style experiments which means the evolution of the electronic
transition rates during coherent excitations are recorded by
measurement of the relaxation of the respective observable after
the excitation as a function of the excitation length $\tau$. This
allows the observation of Rabi oscillation~\cite{Glasbeek:1979}
and also, by application of pulse trains with alternating
excitation phases, the observation of rotary
echoes~\cite{Har:1973,Boe7}. The information gained from these
experiments are coherence times, dephasing times as well as
insights into the coupling between the spin centers involved. For
pESR, the behavior of pairs of spin $s=\frac{1}{2}$ in transient
nutation experiments has been established more than a decade ago
when theoretical studies described the pair evolution in presence
of spin exchange and dipolar coupling~\cite{Gier:1991} and also
under the influence of the hyperfine coupling due to the presence
of nuclei with non--vanishing spin~\cite{web:1997}. These pESR
approaches are similar but not completely applicable to pEDMR and
pODMR experiments due to the difference of their observables.
While the free induction decay of pESR experiments always
represents the polarization of a spin ensemble at the end of the
spin excitation, pODMR/pEDMR transients reflect the permutation
symmetry or antisymmetry of the spin pairs that control the
measured electronic transition rates. The consequences of this
difference can be drastic as explained in detail in a recent study
of spin--dependent transport and recombination through weakly spin
coupled pairs~\cite{Raj:2006}. When excitation intensities are
increased, the Rabi--nutation frequency observed with pEDMR/pODMR
will abruptly double while the nutation observed with pESR will
not. This raises the question of whether the different behavior of
pairs of spin $s=\frac{1}{2}$ as observed for weakly exchange
coupled spin pairs does also imply that pEDMR/pODMR experiments
will differ from the predictions for pESR experiments for strongly
exchange coupled pairs which would render the transient nutation
literature on radical pairs~\cite{Gier:1991,web:1997} not fully
applicable. Moreover, for previous studies for the calculation of
pESR detected transient nutation~\cite{Gier:1991} experiments, one
of the assumptions made was that exchange coupling was smaller
than the separation of the Larmor frequencies within the coupled
radical pairs. This assumption, which is realistic for many
radical pair systems simplified the problem sufficiently enough so
that the results could be calculated analytically. In contrast,
for charge carrier systems in semiconductors with weak
spin--orbital coupling, this assumption does not necessarily
always hold and thus, it limits the applicability of these
previous studies even further.

\begin{figure}
\includegraphics[width=80mm]{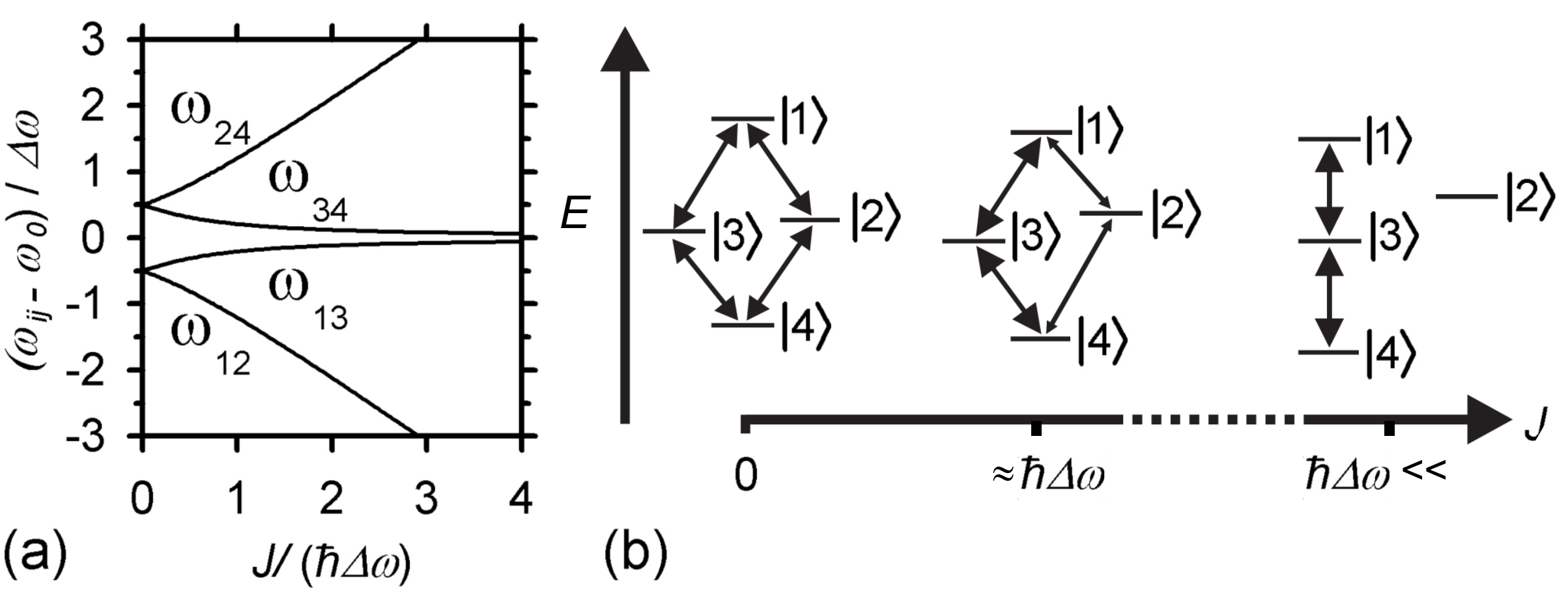}
\caption{(a) Plot of the ESR allowed transition frequencies
$\omega_{ij}=\frac{1}{\hbar}\left(E_i-E_j\right)$ with $\left(ij
\right)\in\left\{12, 13, 24, 34 \right\}$ as function of $J$.
Transitions $\omega_{12}$ and  $\omega_{24}$ involve the state
$|2\rangle$ with high singlet content. For the plot, Larmor
frequencies of $\omega_{a,b}=10\pm0.01\mathrm{GHz}$ have been
assumed. (b) The energy term scheme and ESR allowed transitions
for a spin $\frac{1}{2}$ pair with presence of weak
($\Delta\omega\gg \frac{J}{\hbar}$, left), intermediate
($\Delta\omega\approx \frac{J}{\hbar}$, center) and strong
($\Delta\omega\ll \frac{J}{\hbar}$, right) exchange coupling. The
sketches define the nomenclature of the states and transitions
discussed in this study.} \label{transitions}
\end{figure}

In the following, we present a numerical study of transport and
recombination through localized exchange coupled spin pairs in
semiconductors during coherent spin excitation. The purpose is to
elucidate differences between transient nutation experiments on
exchange coupled radical pairs detected by pESR in contrast to the
detection by pODMR/pEDMR. We focus in particular on the question
when the observed spin pairs act as two single spin systems or one
two--spin systems which turned out to be one of the key
differences for pESR and pEDMR/pODMR detected transient nutation
of weakly coupled spin pairs. Another question that is discussed
is whether magnetic resonance induced triplet--singlet transitions
which become increasingly forbidden with increasing exchange
coupling will reduce the observed signal intensities in the same
way as for pESR or not. It shall be pointed out here that the
focus of this study deals solely with spin--selection rule based
electronic transitions between paramagnetic states in weakly
spin--orbital coupled systems. Typical examples for this would be
charge carrier pairs in semiconductors such as electron--hole
pairs, defect pairs such as donor--acceptor pairs or radical pairs
in molecular systems or solid state host environments. The study
deals with the opposite situation of the pair states discussed in
Ref.~\cite{Raj:2006} which described spin pairs of negligible
exchange. It shall be emphasized that other spin to charge
conversion mechanisms are known (e.g. the nuclear spin measurement
by means of hyperfine coupling to unequally populated quantum Hall
edge channels~\cite{Mach:2003}) whose behaviors during pEDMR
experiments do not necessarily follow the descriptions given here.

\section{Pair model}
The pair model that we use in this study has been described in
detail elsewhere~\cite{Boe:2006,Raj:2006} and therefore it is
described here only briefly. The underlying idea is that spin
dependency of electronic transitions comes from the formation of
spin $s=\frac{1}{2}$ pairs consisting of electrons or holes which
exist for a short time (intermediate pairs) before they either
dissociate or collapse into one doubly occupied electronic state
under formation of a singlet state. The probability for this
collapse depends on the singlet content of the spin pairs when
spin conservation due to sufficiently weak spin--orbit coupling is
given. Thus, the spin dynamics that is determined by the spin
Hamiltonian of these pairs will ultimately control the electronic
transition rates which in turn may determine macroscopic
observables such as luminescence or conductivity. Under magnetic
resonant conditions, a constant magnetic field $\vec
B_0=B_0\mathbf{\hat z}$ and an oscillating magnetic field $\vec
B_1=B_1\left(\mathbf{\hat x+i\hat y}\right)e^{-i\omega t}$ that
rotates in a plane perpendicular to the $B_0$ field, are present.
This leads to a Hamiltonian $\hat H=\hat H_0+\hat H_J+\hat H_1(t)$
of the intermediate spin pairs consisting of the contributions
$\hat
H_0=\frac{1}{2}\mu_BB_0\left[g_a\hat\sigma^a_z+g_b\hat\sigma^b_z\right]$
of the constant magnetic field, $\hat
H_1(t)=\frac{1}{2}g\mu_BB_1\left[\hat\sigma^a_++\hat\sigma^b_+\right]e^{-i\omega
t}$ of the rotating magnetic field and the Heisenberg exchange
coupling $\hat H_J=-J\vec S_a\vec S_b$. In these terms, $\mu_B$
represents Bohr's magneton, $\hat\sigma^i_j$ is the Pauli matrix
$j$ of spin $i$, (with $j\in\{\mathbf{\hat x, \hat y, \hat z}\}$
and $i\in\{a,b\}$), $g_i$ and $\vec S_i$ are the Land\'e--factors
and the spin operators of spin $i$, respectively while $g$
represents the vacuum electron Land\'e factor that is used when
the differences between the weakly coupled spin partners are
negligible (e.g. for the weak influence of the $B_1$ field when
$B_1\ll B_0$). Note that in contrast to previous studies on weakly
coupled spin pairs~\cite{Boe6,Raj:2006}, the pair Hamiltonian here
includes a non--negligible contribution $\hat H_J$ of the exchange
within the pairs. It is the investigation of this coupling which
is in the focus of this study. The time-independent Hamiltonian
$\hat H_0+\hat H_J$ has the eigenvalues \bgan
    E_{1,4}(J) &=& -\frac{1}{4}J\pm\frac{1}{2}\hbar \omega_0 \label{energylevels}\\
    E_{2,3}(J) &=& \frac{1}{4}J\pm\frac{1}{2}\sqrt{J^2 + \hbar^2 {\Delta \omega}^2} \nonumber
\egan where $\omega_0 = \omega_a+\omega_b$ and $\Delta \omega
=\omega_a-\omega_b$ are the sum and the difference of the Larmor
frequencies $\omega_{a,b}$ within the pair. As expected for spin
$s=\frac{1}{2}$ pairs, the energy eigenvalues represent a
four--level system and, as long as the first order processes are
considered only, there are four allowed transitions as well known
from conventional ESR spectroscopy. Fig.~\ref{transitions} shows
the transition frequencies of all four transitions as a function
of the exchange coupling strength $J$. One can see that with
increasing exchange, the energies of the $|1\rangle
\leftrightsquigarrow |3\rangle$ and the $|3\rangle
\leftrightsquigarrow |4\rangle$ transitions (which are the
transitions between the triplet states) will gradually attain the
same value, namely the average $\frac{\omega_0}{2}$ of the Larmor
frequencies whereas the energies of the two transitions involving
the singlet state $|2\rangle$ will become proportional to $J$. In
the case of strong coupling ($\frac{J}{\hbar}\gg\Delta\omega$),
the transition strength into the singlet states will eventually
vanish which is why ESR spectroscopy of strongly coupled pairs
typically is triplet spectroscopy. Since we are concerned with
spin--dependent transport and recombination rates due to the
spin--motion of ensembles of spin pairs, we use in analogy to the
approach of Refs.~\cite{Gier:1991,web:1997,Boe6,Raj:2006} a
density operator $\hat\rho=\hat\rho(t)$ to represent the ensemble
state. In the case of negligible incoherence which means on time
scales faster than the electronic transition times and spin
relaxation times, the dynamics of the ensemble is described by the
Liouville equation
$\partial_t\hat\rho=\frac{i}{\hbar}\left[\hat\rho,\hat
H\right]^-$. When incoherence becomes non--negligible, the
influence of spontaneous electronic transitions such as the
recombination or dissociation of the electronic states as well as
spin relaxation processes must be taken into account by means of
statistical terms $\mathcal{S[\hat\rho]}$ and
$\mathcal{R}[\hat\rho-\hat\rho_0]$, respectively, as it has been
described and outlined for Eq. 1 of Ref.~\cite{Boe6}.

\subsection{PEDMR/pODMR observables}
In contrast to observables of pESR signals which are always
represented by polarization operators applied to a spin ensemble,
the observables for spin--dependent transport and recombination
rates are triplet and singlet operators~\cite{Boe6}. The goal of
this study is to make predictions for pEDMR/pODMR signals as they
are observed in a transient nutation style experiment. Transient
nutation can be observed electrically by application of a short
coherent pulse with length $\tau$ (typically a few nanoseconds)
and the subsequent measurement of spin--dependent transport or
recombination rates through transient measurement of the sample
conductivity. When the applied radiation pulse changes the spin
states of trapped charge carriers resonantly, spin-dependent
transition rates change abruptly and then, after the pulse, they
slowly (on microsecond to millisecond time scales) relax back into
their steady states following an exponential rate transient
\begin{equation}
R(t)=\sum_{i=1}^4\delta\rho_{ii}\left(\tau\right)r_ie^{-r_it}
\label{rate}
\end{equation}
due to the spontaneous spin--dependent electronic transitions
which collapse the coherently excited spin pairs~\cite{Boe:2006}.
In Eq.~\ref{rate},
$\delta\rho_{ii}\left(\tau\right)=\rho_{ii}\left(\tau\right)-\rho_{ii}^S$
whereas $\rho_{ii}\left(\tau\right)$ and $\rho^S_{ii}$ denote the
diagonal elements of the density matrices at the end of the
coherent excitation pulse $\rho\left(\tau\right)$ and the steady
state $\rho^S$, respectively. Both $\rho\left(\tau\right)$ and
$\rho^S$ represent the density operator $\hat\rho$ for the base of
energy eigenstates. The variables $r_i $ denote the
spin--dependent recombination or transport rate coefficients of
the spin states $|i\rangle$ which in presence of exchange coupling
attain the form
\begin{eqnarray}
r_{1,4}&=&r_T \\ \nonumber
r_{2,3}&=&\frac{r_T}{2}\left[1\mp\frac{J}{\hbar\omega_\Delta}\right]
+\frac{r_S}{2}\left[1\pm\frac{J}{\hbar\omega_\Delta}\right]
\label{tranprop2}
\end{eqnarray}
with $\omega_\Delta=\sqrt{J^2/\hbar^2+\Delta\omega^2/4}$ and $r_T$
and $r_S$ denoting transition probabilities of pairs in pure
triplet and pure singlet states, respectively~\cite{Boe6}. Note
that Eq.~\ref{rate} derives from Eq. 17 of Ref.~\cite{Boe6} for
the case of spin--dependent recombination and Eq. 5.17 of
Ref.~\cite{Boe:2006} for the case of spin--dependent transport
under the assumptions that all spin relaxation processes as well
as spin--independent pair dissociation processes are much slower
than the four spin--dependent transitions ($r_i\gg
d,T_1^{-1},T_2^{-1}$). Because of these assumptions, the
relaxation transient of each diagonal element $\delta\rho_{ii}$
towards the steady state becomes a single exponential decay with
decay constant $r_i$.

Similar to the integration of a free induction decay in a pESR
experiment, the integration of the rate relaxation transient in a
pODMR/pEDMR experiment
\begin{equation}
Q(\tau):=\int_0^{t_0} R(t)dt
=\sum_{i=1}^4\delta\rho_{ii}\left(\tau\right)(1-e^{-r_it_0})
\label{rate2}
\end{equation}
contains information about the state of the pair ensemble at $t=0$
which is the moment when the rate relaxation begins at the end of
the excitation pulse with length $\tau$. Note that since $Q(\tau)$
is the time integration of a rate, it is a dimension free variable
representing a number of transitions that take place due to the
pulse excitation. For a pODMR measurement this number translates
directly into a photon number. For pEDMR where a current transient
is integrated into a charge, $Q(\tau)$ represents a number of
elementary charges $e$. The assumption that coherence is preserved
during the excitation pulse implies that
Tr$(\hat\rho(\tau))$=Tr$(\hat\rho^S)$ and, therefore, if we
integrate over large time scales ($t_0\rightarrow\infty$),
$Q(\tau)$ will vanish. This result is reasonable since we
implicitly have assumed for our system that the generation rate of
the spin pairs is constant and not changed by the pulse
excitation. Hence, no matter what time dependence $R(t)$ follows,
averaged over a long time it must assume the generation rate of
intermediate spin pairs. Note that the assumption of a constant
spin pair generation rate is reasonable as long as the
spin--dependent processes change the overall transport or
recombination rates only slightly (less than $10^{-2}$ as it is
the case for most known pEDMR/pODMR signals) since then,
generation rate changes will be negligible second order
effects~\cite{Boe:2006}. Hence, experimentally, $Q(\tau)$ must be
recorded for a finite value of the integration time $t_0$.
Typically it is recorded for $r_{2,3}^{-1}\ll t_0\ll
r_{1,4}^{-1}=r_T^{-1}$ since then, Eq.~\ref{rate2} assumes the
form
$Q(\tau)=\delta\rho_{22}\left(\tau\right)+\delta\rho_{33}\left(\tau\right)$.
For the case of negligible spin--spin coupling within the pairs,
the situation becomes particularly simple since then
$\delta\rho_{22}=\delta\rho_{33}=-\delta\rho_{11}=-\delta\rho_{44}$
due to the symmetry of the pair Hamiltonian and, therefore,
$Q(\tau)\propto\delta\rho_{11}$~\cite{Raj:2006}. In this situation
one may even measure $Q(\tau)$ by integration over the absolute
value of the current change in order to optimize signal strength
as indicated in Fig. 5.7 of Ref.~\cite{Boe:2006}. In contrast,
with increasing exchange $J$, $\delta\rho_{22}\neq\delta\rho_{33}$
and as $J$ becomes very large, $r_2$ approaches $r_T$ which means
that there will be no well defined $t_0$ which fulfills
$r_{1,4}^{-1}\gg t_0\gg r_{2,3}^{-1}$. One solution for this
problem is to set $t_0$ such that $r_{1,3,4}^{-1}\gg t_0\gg
r_{2}^{-1}$ in order to observe transitions into singlet states
only. However, with increasing $J$ the singlet transition
probability diminishes quickly and along with it any measurable
signal. Alternatively, one can set $t_0$ to an arbitrary but well
defined value such as $t_0:=4r_3^{-1}$. This ensures that all
contributions from the singlet transitions and almost all
($1-e^{-4}\approx98\%$) contributions of the triplet signal are
recorded even though both signals will vanish at large $J$ as
well. For the simulations presented in the following we used the
latter assumption ($t_0:=4r_3^{-1}$) for the calculation of the
pEDMR/pODMR observable given in Eq.~\ref{rate2} in order to
investigate both, the influences of singlet and triplet
transitions on pEDMR/pODMR measurements and also, in order to
elucidate how the magnitude of the two signals evolve relatively
to each other as both become smaller with increasing $J$.

\section{Calculations and results}
The behavior of the observable $Q(\tau)$ and its dependence on the
exchange coupling parameter $J$, the frequency $\omega$ and the
amplitude $B_1$ of the exciting radiation, Larmor frequency of the
pair partners $\omega_a$ and $\omega_b$, dissociation rate
coefficient $d$, singlet and triplet recombination rates $r_S$ and
$r_T$ and the generation rate $G$ have been studied. As the field
strength $B_1$ is the parameter that is compared to the system
parameter $\Delta \omega$, we consider three cases for the
strength of the interaction between the $B_1$-field and the
system, namely (i) high power $\gamma B_1/\Delta\omega\gg 1$, (ii)
intermediate power $\gamma B_1/\Delta \omega\approx 1$ and (iii)
low power $\gamma B_1/\Delta\omega\ll 1$ excitation with $\gamma$
denoting the gyromagnetic ratio. Each of these cases is then
simulated for (a) weak ($\Delta\omega\gg \frac{J}{\hbar}$), (b)
intermediate ($\Delta\omega\approx \frac{J}{\hbar}$), and (c)
strong ($\Delta\omega\ll \frac{J}{\hbar}$) exchange coupling. The
evolution of $Q(\tau)$ with increasing pulse length $\tau$ was
calculated by numerical solutions of a systems of 16 partial
differential equations which result from the statistical Liouville
equation
\begin{equation}
\partial_t\hat\rho = \frac{i}{\hbar}\left[\hat\rho,H\right]^- + \mathcal{S}[\hat\rho]+\mathcal{R}[\hat\rho-\hat\rho_0]
\label{liouv}
\end{equation}
in which the spin relaxation processes
$\mathcal{R}[\hat\rho-\hat\rho_0]\approx 0$ were neglected but not
the incoherence $\mathcal{S}[\hat\rho]\neq 0$ induced by the
\onecolumngrid
\begin{figure*}
\includegraphics[width=170mm]{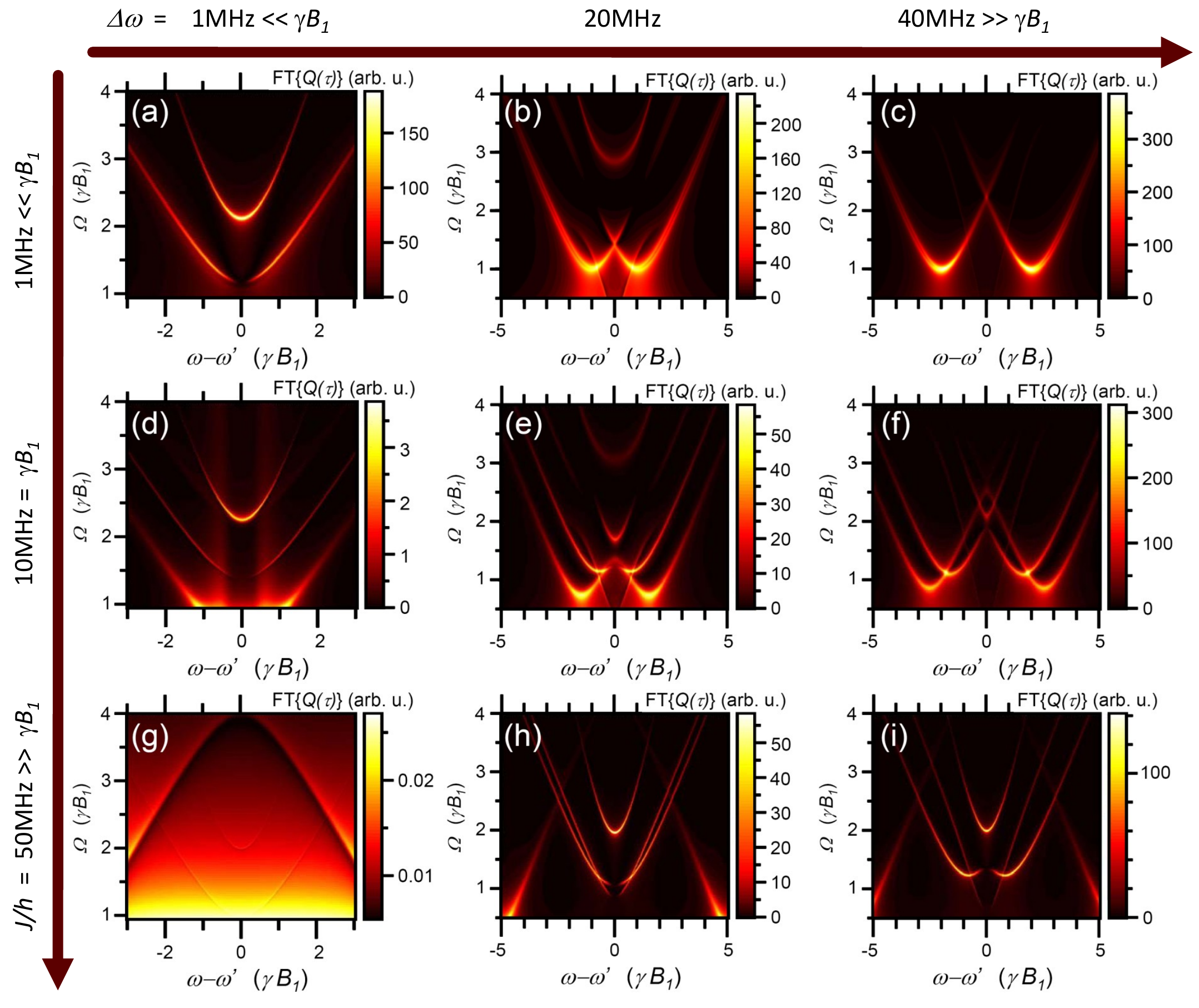}
\caption{Plots of the Fourier transform FT$\{Q(\tau)\}$ of the
observable $Q(\tau)$. The data displays the Fourier components of
$Q(\tau)$ as a function of the excitation (driving) frequency,
$\omega$. Data sets are represented in arbitrary but equal units.
They were simulated with Larmor separations of $\Delta\omega=1$MHz
(plots (a),(d),(g), left column), $\Delta\omega=20$MHz (plots
(b),(e),(h), center column) and $\Delta\omega=40$MHz (plots
(c),(f),(i), right column) as well as exchange coupling strengths
with in the pairs of $J/h=1$MHz (plots (a),(b),(c), first row),
$J/h=10$MHz (plots (d),(e),(f), second row) and $J/h=50$MHz (plots
(g),(h),(i), third row).} \label{fig2}
\end{figure*}
\twocolumngrid
spin--dependent electronic transitions. The time domain signal
$Q(\tau)$ was calculated for pulse length $0\leq\tau\leq1.5\mu$s
with a resolution of 1000 steps per nanosecond. These transients
were then Fourier transformed in order to reveal the frequency
components $FT\{Q(\Omega)\}$. In order to test the qualitative
correctness of these results, an independent calculation of the
nutation frequencies of the Hamiltonian $\hat H=\hat H_0+\hat
H_J+\hat H_1(t)$ was conducted for each used set of parameters and
compared to the results obtained numerically. In these test
calculations the time evolution of the spin ensemble was studied
by simply applying the time evolution operator
$\exp\left(-\frac{i\hat H^*\tau}{\hbar}\right)$ (with $\hat H^*$
representing the time independent Hamiltonian of the pair in a
reference frame that rotates with the circularly polarized
magnetic field of the exciting radiation) to the reference frame
independent initial state defined in Ref~\cite{Raj:2006}. The test
confirmed our numerical results. In Fig.\ref{fig2}(a) to (i) the
results of the numerical solutions of the stochastic Liouville
equation (Eq.~\ref{liouv}) are presented. They display the
intensity of the pEDMR/pODMR signal as a function of their
frequency components $\Omega$ and the frequency $\omega$ of the
applied driving field. Note that these frequencies are expressed
in terms of the field strength $\gamma B_1$ of the driving field.
The parameters used for the calculation of the data displayed in
Fig.~\ref{fig2} represent the case of $B_1\ll B_0$ and thus
$\Omega\ll \omega$ as found in most pESR experiments. In fact the
parameters used here represent typical values that can be
established in modern, commercially available pESR X-Band
spectrometers: For all data sets $\omega^\prime =
\frac{\omega_a+\omega_b}{2} =10\mathrm{GHz}$, $\frac{\gamma}{2\pi}
B_1\,=\,10\mathrm{MHz}$ and
$\frac{\omega_0}{2\pi}\,=\,10\mathrm{GHz}$. Furthermore, for all
simulations we assumed that the signal was caused by a
spin-dependent electronic transition based on the intermediate
pair model described above whose intermediate pairs had a singlet
recombination rate coefficient $r_S=10^{6}s^{-1})$, a triplet
recombination rate coefficient $r_T=10^{4}s^{-1})$, and a pair
dissociation rate coefficient $d=10^{3}s^{-1})$ that is much
smaller than the recombination probabilities. Note that the orders
of magnitude of these values correspond to experimental data
obtained from spin-dependent recombination processes at defects in
different silicon morphologies~\cite{Boe7}. The generation rate of
the simulated pair ensemble was chosen arbitrary since it only
scales the intensity of the calculated observable. However note
that this arbitrary value was kept constant for all simulations
presented here in order to make the signal intensities as plotted
in Fig.~\ref{fig2} comparable. The data sets displayed in
Fig.~\ref{fig2} represent the combination of three different
Larmor separations ($\Delta \omega$=1, 20, and 40MHz) and three
different exchange coupling constants ($J/h$=1, 10 and 50MHz). The
choice of these values was made in order to establish the
qualitative behavior of the system when $\Delta\omega, J/\hbar\ll
\gamma B_1$, $\Delta\omega \ll \gamma B_1 \ll J/\hbar$,
$J/\hbar\ll \gamma B_1 \ll \Delta\omega$ and $\gamma
B_1\ll\Delta\omega, J/\hbar$ as well as the intermediate cases
where two or all three of these three parameters have comparable
magnitudes. Note that for all cases it is assumed that
$J/\hbar,\Delta\omega,\gamma B_1\ll \omega_a,\omega_b$ which
implies that the results presented here will not be applicable to
very strongly coupled excitonic states
($J/\hbar\gg\omega_a,\omega_b$) that can be found in many polymers
and also in quantum dots. Note however that we will find in the
following for cases of strongly yet not very strongly coupled
pairs ($\omega_a,\omega_b\gg J/\hbar\gg\Delta\omega$) that pEDMR
and pODMR signals will vanish and hence, the consideration of very
strongly coupled systems is not relevant in this case.

\section{Discussion}
The simulated data displayed in figs.~\ref{fig2}(a-f) shows
different hyperbola shaped structures which are known and expected
for a resonant system with different eigenfrequencies. The
symmetry centers of these structures represent the transition
frequencies $\omega_i$ which correspond for all displayed data
sets (and their corresponding parameter sets) to the energy
differences that can be derived from Eq.\ref{energylevels}.
Similarly as for the weakly exchange--coupled case described in
Ref.~\cite{Raj:2006}, the hyperbola shapes are caused by the
increase of the transient nutation frequencies as the excitation
frequency is detuned from a given resonance as described by Rabi's
formula $\Omega_i=\sqrt{(\gamma B_1)^2+(\omega_i-\omega)^2}$. In
contrast to Ref.~\cite{Raj:2006}, we see up to six different
frequency components that exist for any given excitation
frequency. The higher number of nutation components is anticipated
since the introduction of the exchange Hamiltonian removed the
symmetry from the four energy levels leading to four instead of
just two transition energies (see Fig.~\ref{transitions} for the
case of intermediate coupling $J\approx\hbar\Delta\omega$) and
correspondingly, to four instead of two associated nutation
frequencies. With two nutation frequencies given for the weakly
exchange coupled case one can anticipate up to four different
Rabi-frequency components when permutation symmetry and not
polarization is detected with pEDMR and pODMR. Hence, the beat
oscillations of the pair nutations can become dominating under
certain conditions (for weakly coupled pairs this condition was
$\Delta\omega\ll\gamma B_1$). Therefore, with four nutation
frequencies given for the strongly exchange coupled case one can
anticipate an even larger number of Rabi-frequency components
since as many as 12 combinations of these nutation frequencies are
conceivable, even though not all of these may be dominant
contributions. The presence of the beat oscillations can be
verified from Fig.~\ref{fig2}. The hyperbola shaped features
around a resonance $\omega_i$ turn into linear functions in their
far off-resonant regions. The slopes of these linear functions are
proportional to $(\omega_i-\omega)$ for single nutation
components. However, since beat oscillations resemble either the
sum or the difference of single nutation components, their
frequencies are either independent of $\omega$ or proportional to
$2(\omega_i-\omega)$. A comparison of the resonance hyperbolas in
the different graphs of Fig.~\ref{fig2} shows that the beat
oscillations are present whenever either $\gamma
B_1\gg\Delta\omega$ (Fig.~\ref{fig2} a, d, g) or
$\frac{J}{\hbar}\gg\Delta\omega$ (Fig.~\ref{fig2} g, h, i) and in
fact the only data set which did not indicate any beat component
at all is shown in plot Fig.~\ref{fig2}, c where $\gamma B_1,
\frac{J}{\hbar} \ll\Delta\omega$. This confirms that when
permutation symmetry is the observable (meaning for pEDMR or pODMR
measurements) it is either the exchange coupling $J$ or the
interaction of both spins with the strong excitation fields
$\gamma B_1$ which determine whether the observed oscillations are
due to nutations of the individual spin pair partners or the beat
oscillations thereof.

The second insight gained from Fig.~\ref{fig2} is the magnitude of
the nutation frequencies as well as their beat oscillations. For
the pESR measurement of transient nutation it had been
shown~\cite{Gier:1991} that for large exchange coupling
($\frac{J}{\hbar}\gg\Delta\omega,\gamma B_1 $), the on--resonance
nutation frequencies associated with the triplet transitions
approach values of $\Omega=\sqrt{2}\gamma B_1$ whereas the
nutation frequencies associated with the singlet transitions
(whose oscillator strengths sharply drop as $J$ increases)
approach $\Omega=0$. In agreement with these predictions, we
obtain for the nutation frequencies of the singlet transitions
values that decrease from $\Omega=\gamma B_1$ to $\Omega=0$ as $J$
increases and for the nutation frequencies of the triplet
transitions values that rise from $\Omega=\gamma B_1$ as $J$
increases. However, as $J$ becomes very large, the triplet
nutation frequencies become less and less dominant while beat
oscillations become more dominant. As one can see from
Fig.~\ref{fig2}, these beat oscillations exhibit on-resonance
frequencies of $\Omega=2\gamma B_1$ and not
$\Omega=2\sqrt{2}\gamma B_1$ as one would anticipate from a simple
addition of two frequencies. This behavior is consistent with the
"nutation frequency doubling" described for the pEDMR/pODMR
detection of weakly coupled pairs under strong $B_1$ fields.
However, it is in contrast to the observations with pESR of
strongly exchanged coupled spin pairs and at this time we are not
aware of a straightforward picture that could provide a
qualitative interpretation of this behavior.

The third observation that we obtain from the simulated data is
that the ratio of the signal strengths of the observed singlet
transitions and the triplet transitions does not change
significantly as $J$ is increased: In pESR spectroscopy of
exchange coupled spin pairs, the strong decline of the oscillator
strength of singlet transitions leads to a disappearance of the
singlet signal in comparison to a triplet signal whose oscillator
strength increases. In contrast, for pEDMR/pODMR experiments, the
increase of $J$ causes a drop-off of both the singlet and triplet
signals. The singlet signal drops off for the same reasons as the
singlet strength of pESR signals. In contrast, the triplet signal
drops off since the transition from one triplet state into another
triplet state does not change the permutation symmetry of the
pairs and hence, a change of the detected electronic transition
rate does not take place either. Note that the data sets presented
in Fig.~\ref{fig2} are plotted with different color scales in
order to display the data with optimal contrast. However, the
arbitrary units to which the different color scales translate are
equal for all nine displayed data sets. Thus, Fig.~\ref{fig2}
illustrates how the signals drop as either the exchange coupling
$J$ or the Larmor separation $\Delta\omega$ increases. The
strongest decline of the signal intensity is in fact given in plot
(g) where $\Delta\omega$ is minimized and $J$ is maximized. Note
however that in spite of this decrease of the signal strengths,
the relative strengths of the singlet and triplet contributions
remain of comparable magnitude. Hence, as long as the sensitivity
of a given pEDMR or pODMR experiment allows the detection of any
signal under strong coupling, both the triplet and the singlet
contributions should be equally well observable. This realization
can be important for the verification of the nature of an observed
spin pair system. It may be a way to distinguish experimentally
the difference between a strongly exchange coupled system
($\frac{J}{\hbar}\gg\Delta\omega$) and a weakly exchange coupled
system where $\frac{J}{\hbar}\ll\Delta\omega\ll\gamma B_1$.

\section{Summary and Conclusions}
The dynamics of spin--dependent electronic transport and
recombination rates through strongly exchange--coupled spin pairs
during coherent electron spin resonant excitation as it would be
observed with pEDMR or pODMR detected transient nutation
experiments has been simulated by numerical solution of the
stochastic Liouvolle equation which accounted for incoherent
effects. From the results we conclude that significant qualitative
and quantitative differences can exist between pEDMR/pODMR
experiments in comparison to pESR experiments conducted on the
same systems. Similarly as for the weakly exchange coupled case,
we observe that beat oscillations of the spin nutations and not
the spin nutation alone dominate the transport or recombination
rates whenever either the exchange coupling strength $J$ or the
$B_1$ field exceed the Larmor separation $\Delta\omega$ within the
pair. Moreover, while the intensities of the rate oscillations
decrease with increasing exchange within the spin pairs, the
singlet and triplet signals retain their relative strength. This
means that pEDMR and pOMDR for both of which permutation symmetry
is utilized as observable can provide insights and information
about the nature of the observed spin systems which would hardly
or not at all be accessible with conventional radiation (and,
therefore, polarization-) detected electron spin resonance
spectroscopy.

\begin{acknowledgments}
A. G. gratefully acknowledges support by the Swiss National
Foundation under Grant no. 200020-107428/1. C. M., S. D. B. and F.
G. gratefully acknowledge Financial support of the Deutsche
Forschungsgemeinschaft, of the Fonds der Chemischen Industrie, and
the European Graduate College "Electron--Electron Interactions in
Solids" Marburg--Budapest.
\end{acknowledgments}

\bibliographystyle{prsty}

\pagebreak

\end{document}